\documentclass[3p,times,procedia]{elsarticle}
\usepackage{nupha_ecrc}

\volume{00}

\firstpage{1}

\journalname{Nuclear Physics A}

\runauth{}

\jid{nupha}

\jnltitlelogo{Nuclear Physics A}

\usepackage{amssymb}

\usepackage{lineno}

\usepackage[figuresright]{rotating}

\def\jpsi{$\mathrm{J}/\psi${}}
\def\psip{$\psi(\mathrm{2S})$}

\begin{document}

\begin{frontmatter}



\dochead{XXVIth International Conference on Ultrarelativistic Nucleus-Nucleus Collisions\\ (Quark Matter 2017)}
\runauth{E. L. Kryshen} 

\title{Photoproduction of heavy vector mesons in ultra-peripheral Pb-Pb collisions}


\author[label1]{E. L. Kryshen}
\author{for the ALICE Collaboration}

\address[label1]{Petersburg Nuclear Physics Institute, Gatchina, Russia}

\begin{abstract}
Ultra-peripheral Pb-Pb collisions, in which the two nuclei pass close to each other, but at an impact parameter greater than the sum of their radii, provide information about the initial state of nuclei. In particular, heavy vector meson production, where the particle mass sets a hard scale, proceeds in such collisions by photon-gluon interactions, and gives access to nuclear PDFs. The ALICE collaboration has published measurements of $J/\psi$ and $\psi(2S)$ photoproduction in ultra-peripheral collisions in LHC Run 1 at forward ($J/\psi$) and mid-rapidity, and has obtained a substantially larger data set in 2015 from LHC Run 2, allowing much more detailed studies of the production mechanism to be performed. In particular, the increased energy and more detailed measurements in the forward region in Run 2 give access to significantly lower values of Bjorken-$x$ than in previous studies. In this talk, the latest available results from Run 2 will be given.
\end{abstract}


\end{frontmatter}



\section{Introduction}
\label{intro}

Lead nuclei, accelerated at the LHC, are sources of strong electromagnetic fields, which are equivalent to a flux of quasi-real photons, thus Pb--Pb collisions can be used to measure $\gamma$Pb interactions in a new kinematic regime. These interactions are usually studied in ultra-peripheral collisions (UPC), characterised by impact parameters larger than the sum of the radii of the incoming nuclei, in which hadronic interactions are strongly suppressed~\cite{Baltz:2007kq,Contreras:2015dqa}. Coherent heavy quarkonium photoproduction is of particular interest since, in leading order perturbative QCD, its cross section is proportional to the squared gluon density of the target~\cite{Ryskin:1992ui}. LHC kinematics corresponds to Bjorken-$x$ ranging from $x \sim 10^{-2}$ down to $x \sim 10^{-5}$, while the heavy-quark mass requires a virtuality $Q^2$ larger than a few GeV$^2$, hence introducing a hard scale. Quarkonium photoproduction in \mbox{Pb--Pb} UPC provides a direct tool to study nuclear gluon shadowing effects~\cite{Guzey:2013xba}, which are poorly known and play a crucial role in the initial stages of heavy-ion collisions.

\section{Charmonium photoproduction in Pb-Pb UPC at $\sqrt{s_{\rm NN}} = 5.02$ TeV}
ALICE has previously published results on coherent \jpsi\ and \psip\ photoproduction in Pb--Pb UPC at~$\sqrt{s_{\mathrm{NN}}} = 2.76\ {\rm TeV}$~\cite{Abelev:2012ba,Abbas:2013oua,Adam:2015sia}. This work presents the latest results from Run 2 on J$/\psi$ photoproduction in Pb--Pb UPC at~$\sqrt{s_{\mathrm{NN}}} = 5.02\ {\rm TeV}$.

Charmonium photoproduction was studied both at central and forward rapidity. The forward UPC trigger in Run 2 required two unlike-sign tracks with $p_{\rm T} > 1$\ GeV/$c$ in the muon spectrometer and a veto in the \mbox{V0-A} ($2.8 < \eta <5.1$),~\mbox{AD-A} ($4.9 < \eta < 6.3$) and~\mbox{AD-C} ($-7.0 < \eta < -4.8$) scintillator arrays. (A full description of ALICE can be found in~\cite{Aamodt:2008zz}.) Event emptiness at central rapidity was further ensured by vetoing activity in the silicon-pixel detector (SPD). Events with opposite-sign dimuons in the rapidity range from $-4.0$ to $-2.5$ were selected in the offline analysis. The event sample corresponds to an integrated luminosity of about~$216\ \mu{\rm b}^{-1}$.

\begin{figure}[b!]
\vspace*{-0.5cm}
\centering
\includegraphics[width=0.49\linewidth]{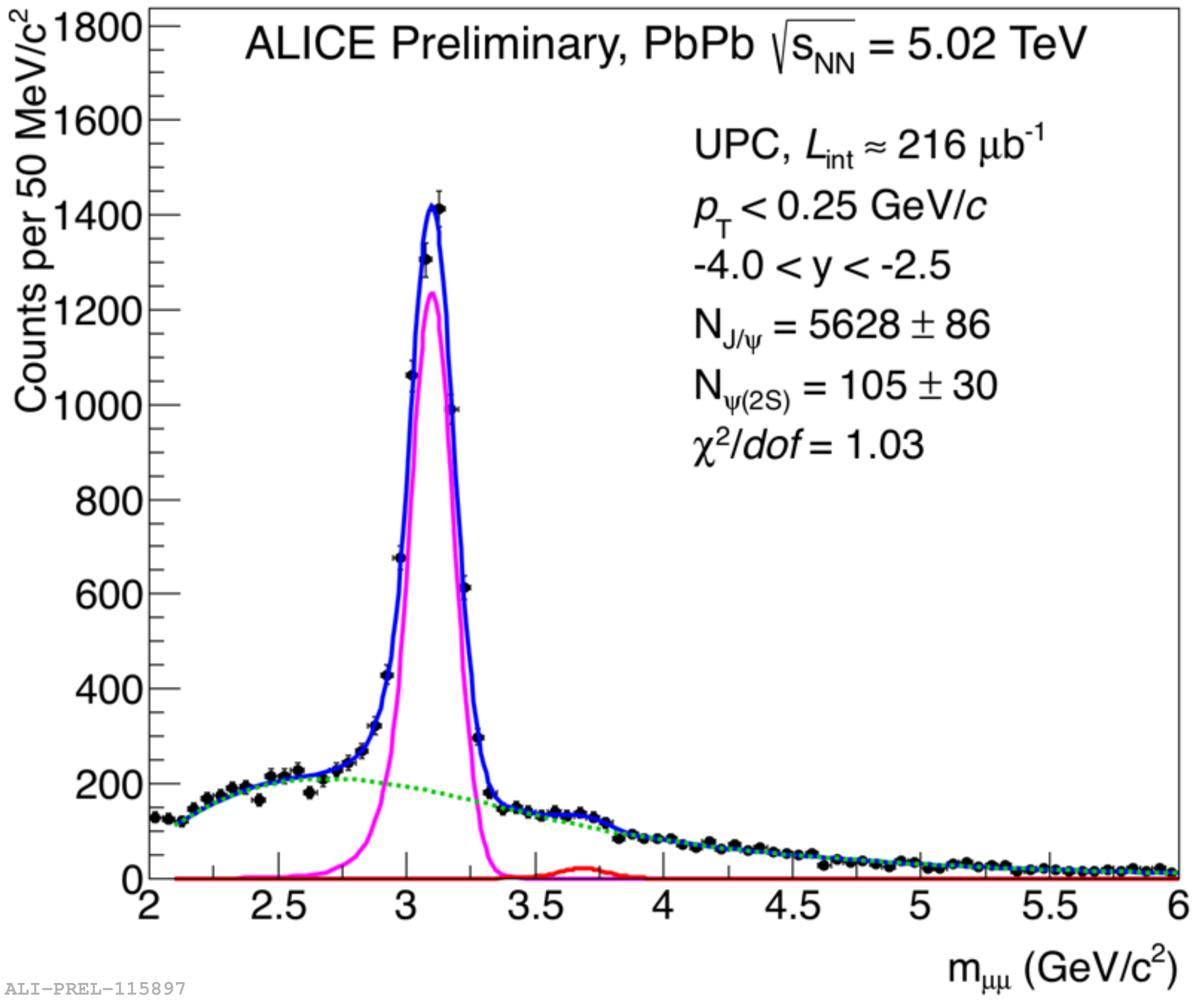}
\includegraphics[width=0.49\linewidth]{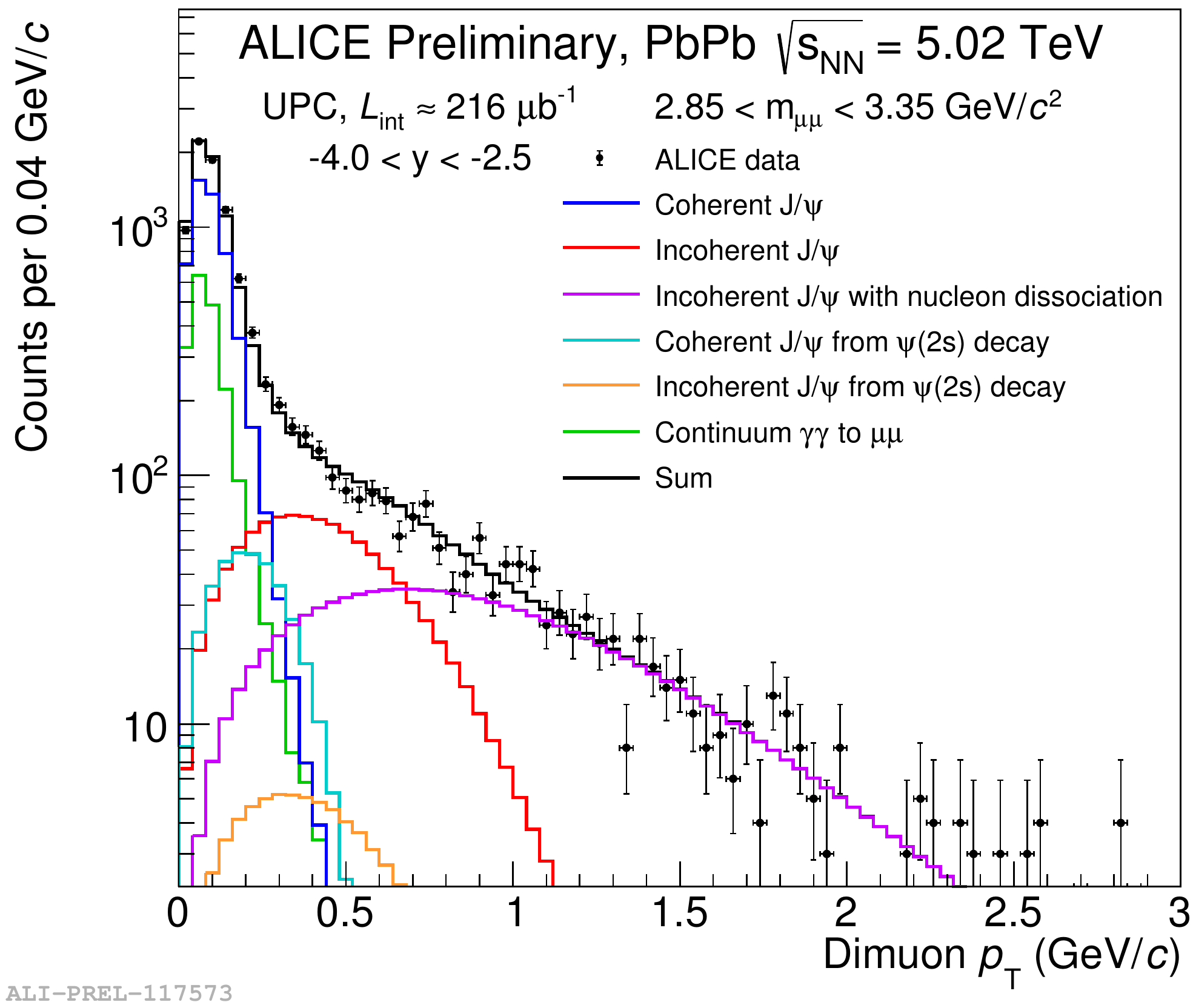}
\label{fig:1}
\caption{Left: invariant mass distribution for unlike-sign dimuons with pair $p_{\rm T} < 0.25 {\rm\  GeV}/c$ and rapidity $-4.0<y<-2.5$  in ultra-peripheral Pb-Pb collisions at $\sqrt{s_{\rm NN}} = 5.02$ TeV. Right: transverse momentum distribution for unlike-sign dimuons around $J/\psi$ mass fitted summing six diffirent Monte Carlo templates.}
\end{figure}

The invariant mass distribution for opposite-sign dimuons with pair transverse momentum $p_{\rm T}$ below 0.25 GeV/$c$ is shown in Fig.~\ref{fig:1}, left. $J/\psi$ and $\psi(2S)$ signals were fitted with Crystal Ball functions on top of a background described by a convolution of an exponential and a trigger turn-on polynomial function. The obtained $J/\psi$ yield is a factor 50 higher compared to Run 1 results at forward rapidity~\cite{Abelev:2012ba} thanks to higher integrated luminosity, improved trigger logic, wider rapidity range and increased beam energy. The $\psi(2S)$ signal can be observed at about 3$\sigma$ significance level. The background shape is in good agreement with continuum $\gamma\gamma \to \mu\mu$ production.

The transverse momentum distribution for dimuons around the $J/\psi$ mass is shown in Fig.~\ref{fig:1}, right. It was fitted with Monte-Carlo templates 
produced using the STARLIGHT event generator~\cite{Klein:2016yzr} and 
corresponding to different production mechanisms. Coherent \jpsi\ photoproduction, when a photon interacts coherently with the~whole nucleus, is characterized by a narrow transverse momentum distribution with~$\langle p_{\rm T} \rangle \sim 60$~MeV$/c$. In the incoherent case the photon couples to a single nucleon. If the target nucleon stays intact, the charmonium $p_{\rm T}$ distribution is driven by the nucleon form factor with $\langle p_{\rm T} \rangle \sim 400$~MeV$/c$. $J/\psi$ photoproduction on a single nucleon can be also accompanied by nucleon dissociation. Dissociative $J/\psi$ photoproduction template 
produced with the H1 parameterization~\cite{Alexa:2013xxa} 
was taken into account to describe high-$p_{\rm T}$ tail. Contributions from continuum dimuon production and feed-down from \psip\ decays were also taken into account in the fits.

\begin{figure}[t]
\centering
\includegraphics[trim={0 1cm 0 0},clip,width=0.56\linewidth]{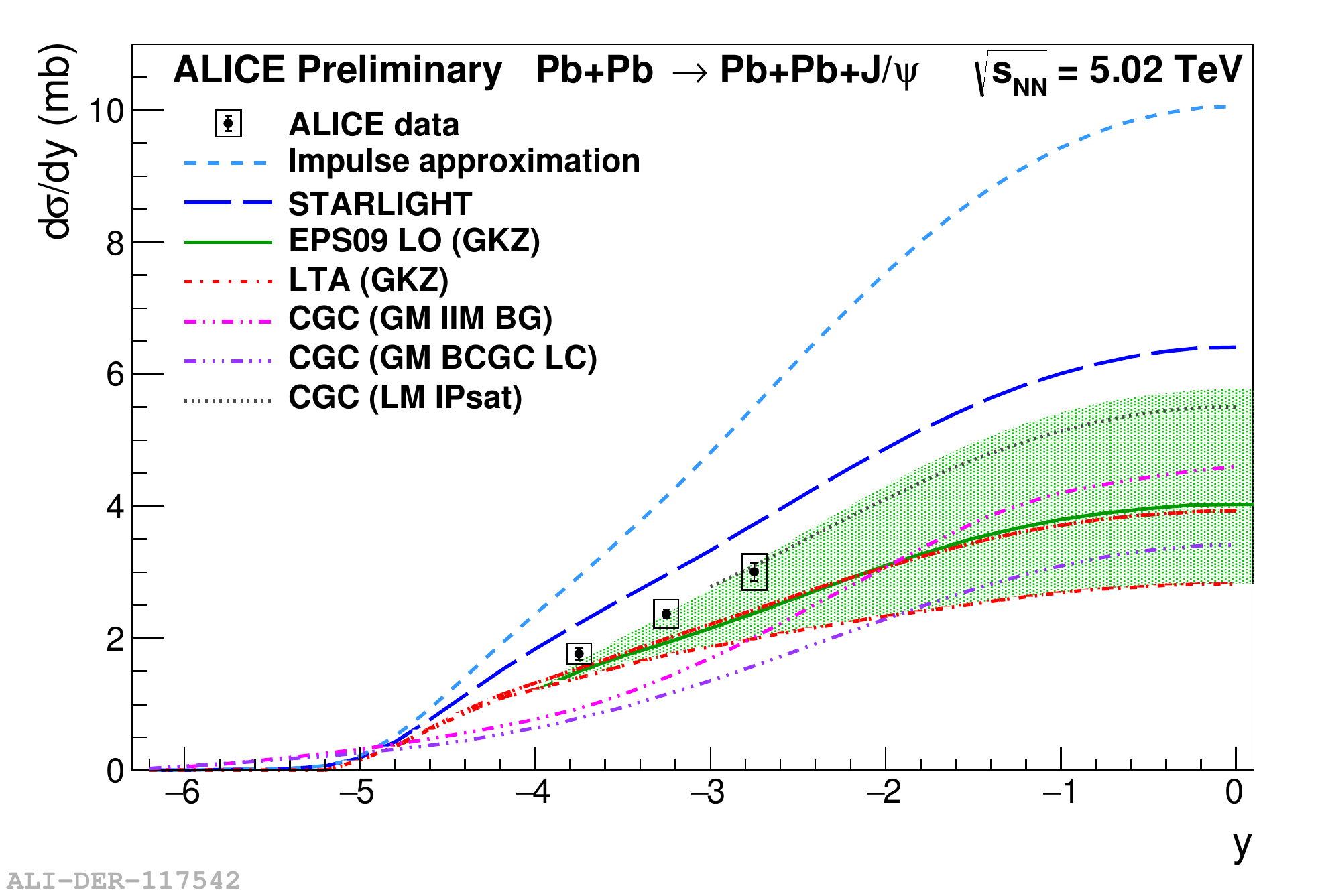}
\caption{
Measured coherent differential cross section of $J/\psi$ photoproduction in ultra-peripheral Pb-Pb collisions at $\sqrt{s_{\rm NN}} = 5.02$ TeV. The error bars correspond to the statistical uncertainties, the open boxes to the systematic uncertainties. Results from various models are also shown.
\label{fig:2}
} 
\end{figure}

The ALICE results on the coherent $J/\psi$ photoproduction cross section at forward rapidity in ultra-peripheral Pb-Pb collisions at $\sqrt{s_{\rm NN}} = 5.02$ TeV are compared to several theoretical calculations  in Fig.~\ref{fig:2}. The impulse approximation, the baseline calculation in the absence of any nuclear effects, and the STAR\-LIGHT event generator~\cite{Klein:2016yzr}, based on the vector dominance model, overpredict the data. Several predictions using the Colour Glass Condensate (CGC) framework under different assumptions have been provided by Gonçalves, Machado et al.~\cite{Goncalves:2014wna,Santos:2014zna} and Lappi and Mantysaari~\cite{Lappi:2013am}. The latter model provides good agreement with the data however its range of validity does not span all the experimental points. Finally, Guzey, Kryshen and Zhalov provide two calculations, one based on the EPS09 framework and the other on the Leading Twist Approximation (LTA)~\cite{Guzey:2016piu}. The LTA curve is lower than the EPS09 one, and underpredicts the data, while the EPS09 mid-value underpredicts the data but remains compatible with it within the model uncertainties.

The central UPC trigger, corresponding to an integrated luminosity of about~$95\ \mu{\rm b}^{-1}$, involved vetoes in the V0 and AD detectors and topological requirements in the SPD and the Time-Of-Flight detector (TOF). This trigger provided factor 5 higher statistics compared to Run 1 results. Tracks were reconstructed in the ALICE central barrel, the energy deposition in the Time Projection Chamber (TPC) was used for the particle identification. 

\begin{figure}[h!]
\centering
\includegraphics[trim={0 0 0 1.5cm},clip,width=0.49\linewidth]{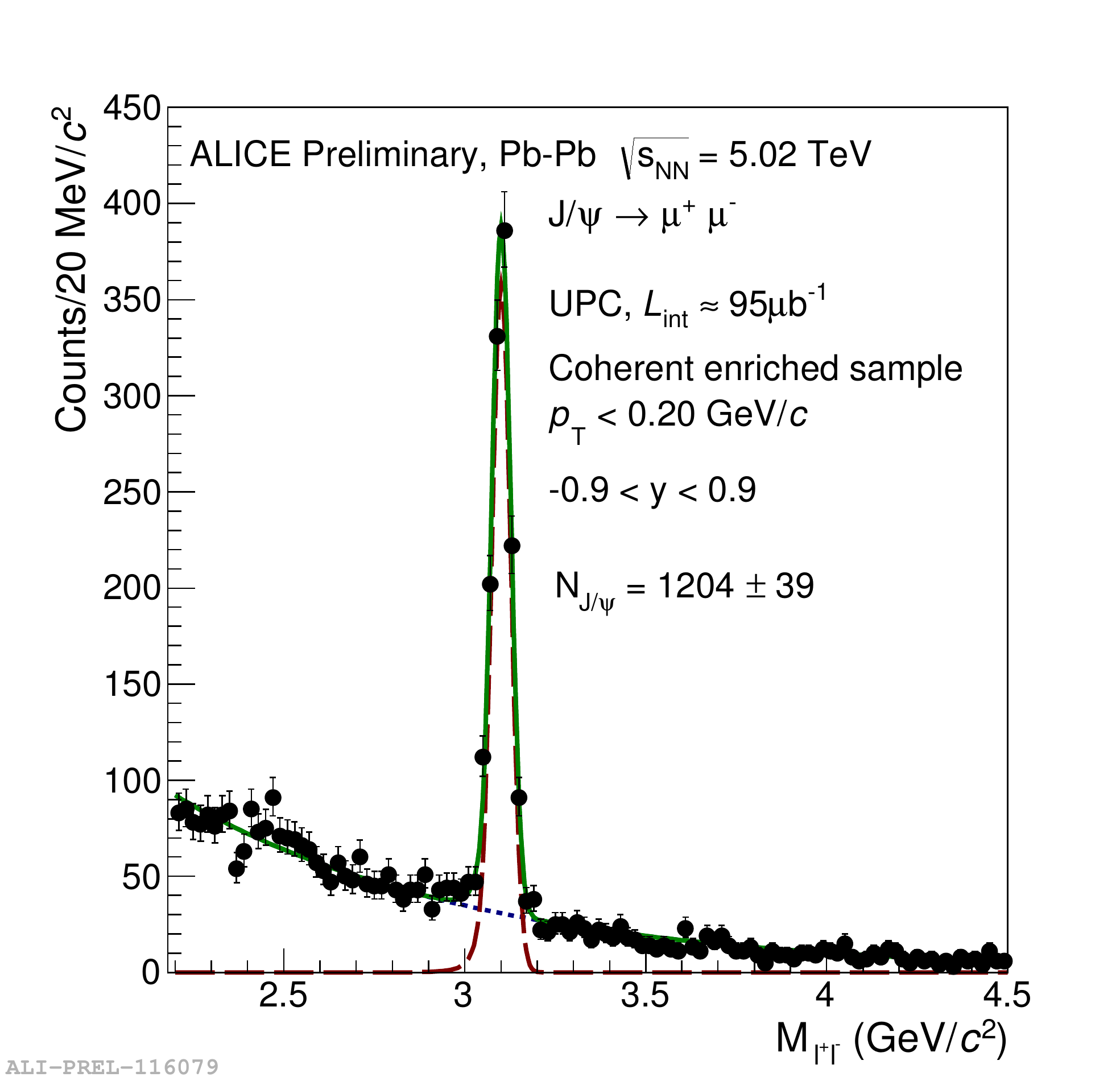}
\includegraphics[trim={0 0 0 1.5cm},clip,width=0.49\linewidth]{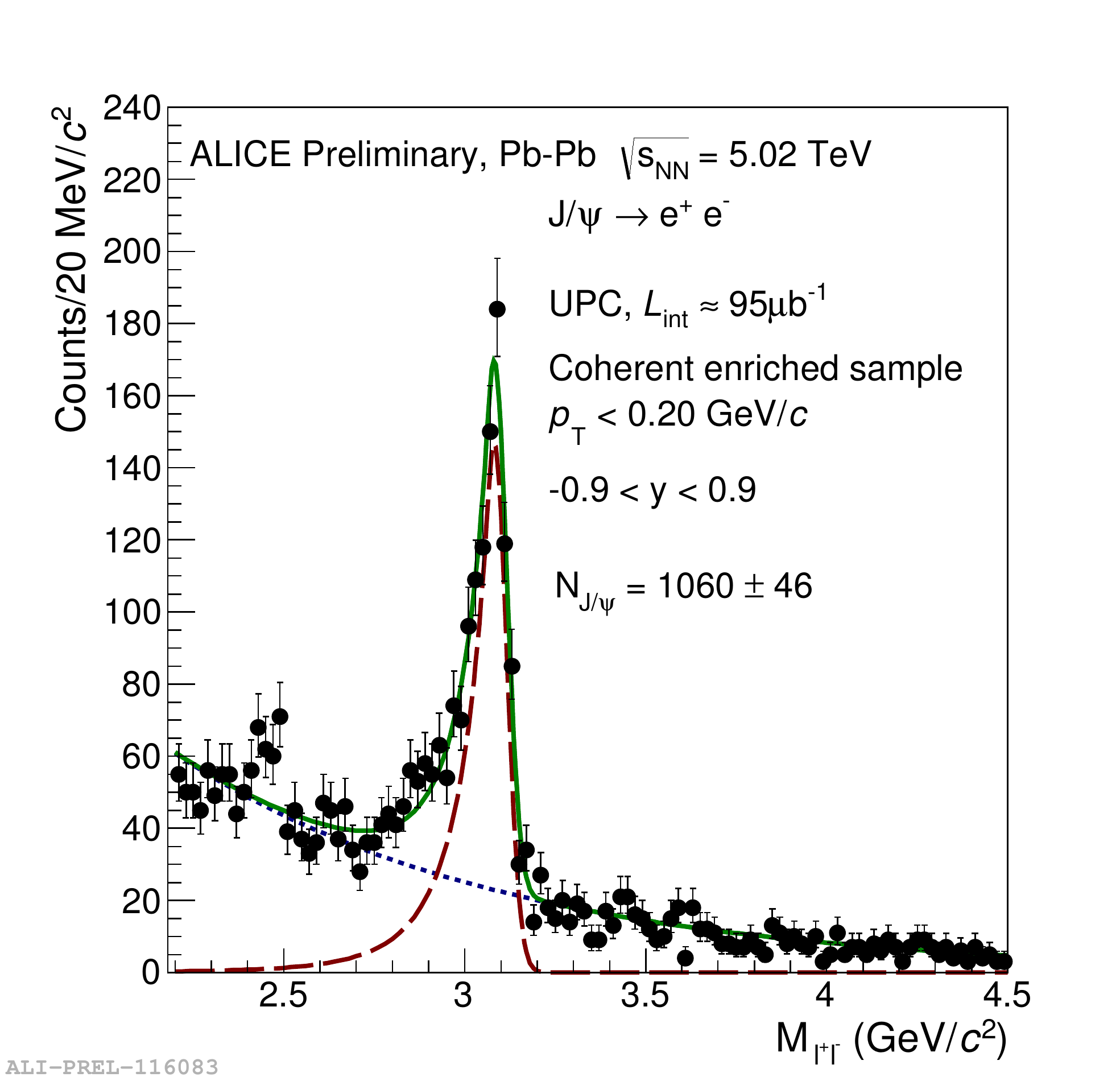}
\caption{
\label{fig:3}
Invariant mass distribution for unlike-sign dimuons (left) and dielectrons (right) with pair $p_{\rm T} < 0.2 {\rm\  GeV}/c$ and rapidity $|y|<0.9$  in ultra-peripheral Pb-Pb collisions at $\sqrt{s_{\rm NN}} = 5.02$ TeV.  The $J/\psi$ signal is fitted with a Crystal Ball function on top of dilepton continuum fitted with an exponential function.}
\end{figure}

\begin{figure}[t]
\centering
\includegraphics[trim={0 0 0 1.25cm},clip,width=0.46\linewidth]{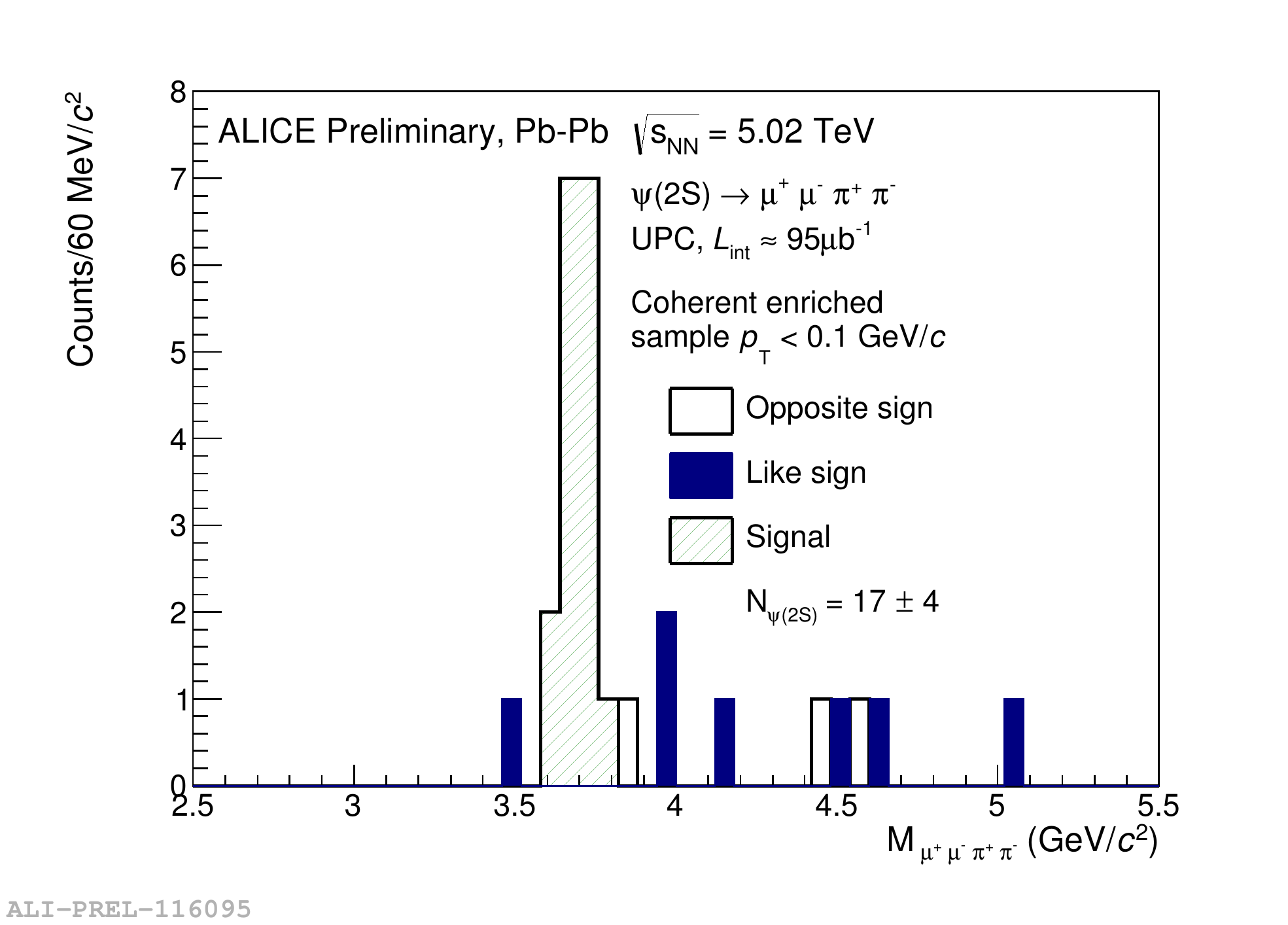}
\includegraphics[trim={0 0 0 1.25cm},clip,width=0.46\linewidth]{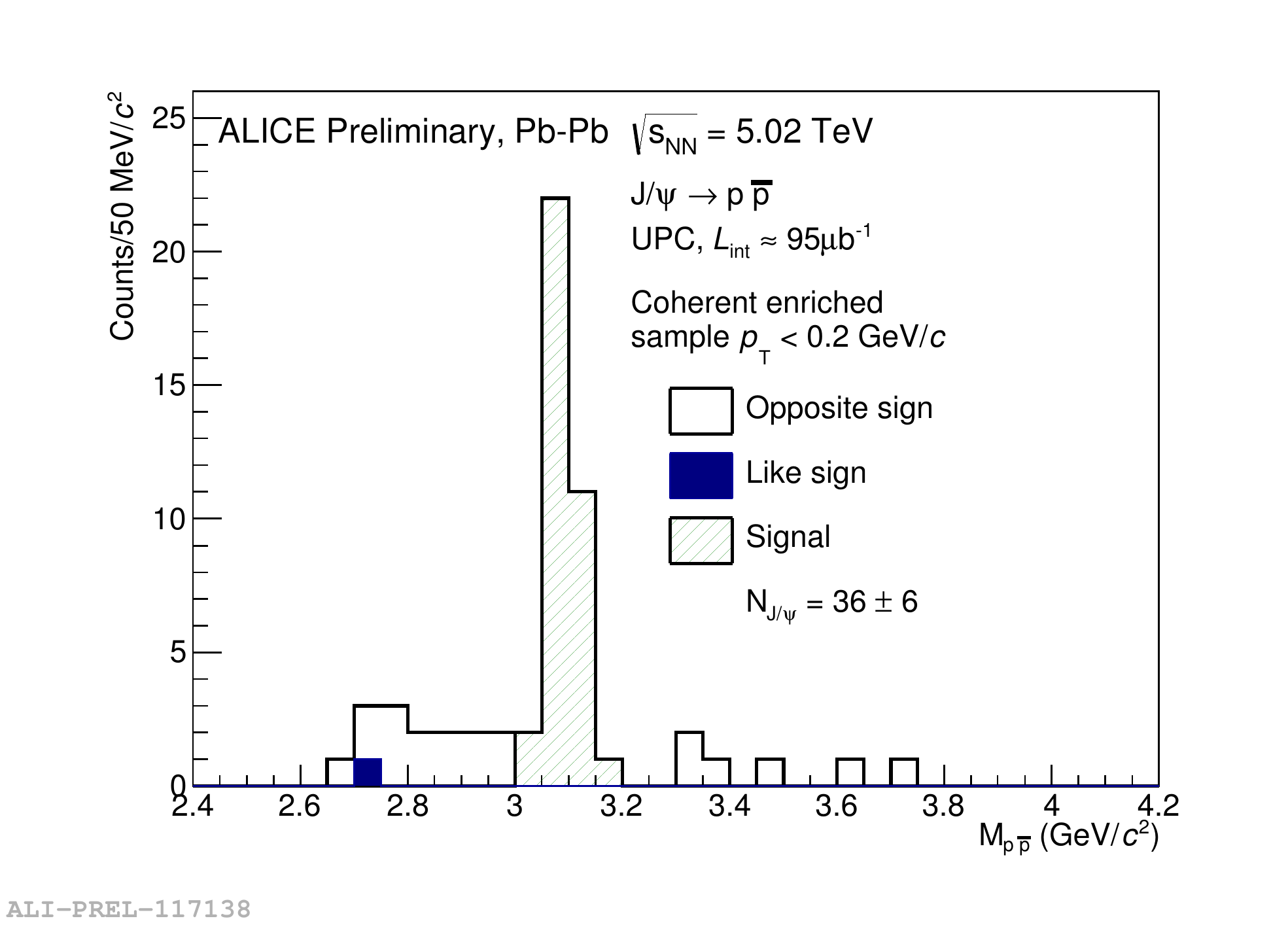}
\caption{
\label{fig:4}
Invariant mass distribution for $\psi(2S) \to \mu^+\mu^-\pi^+\pi^-$ channel (left) and $J/\psi \to p \bar p$ channel (right) at central rapidity in ultra-peripheral Pb-Pb collisions at $\sqrt{s_{\rm NN}} = 5.02$ TeV.}
\end{figure}

Invariant mass distributions for unlike-sign dimuons and dielectrons with pair $p_{\rm T} < 0.2 {\rm\  GeV}/c$ at central rapidity are shown in Fig.~\ref{fig:3}. The coherent $J/\psi$ signal has been also observed in the $p \bar p$ channel, see Fig.~\ref{fig:4}, right. Similar to Run 1~\cite{Adam:2015sia}, the $\psi(2S)$ signal was extracted in the $\psi(2S) \to \mu^+\mu^-\pi^+\pi^-$ and $\psi(2S) \to e^+e^-\pi^+\pi^-$ channels. The invariant mass distribution for $\psi(2S) \to \mu^+\mu^-\pi^+\pi^-$ channel is shown in Fig.~\ref{fig:4}, left. 

UPC results at central rapidity will provide further constraints on the nuclear gluon shadowing at $x \sim 10^{-3}$.

\section{$J/\psi$ photoproduction in p--Pb UPC  at $\sqrt{s_{\rm NN}} = 8.16$ TeV}
\begin{figure}[h]
\centering
\includegraphics[width=0.38\linewidth]{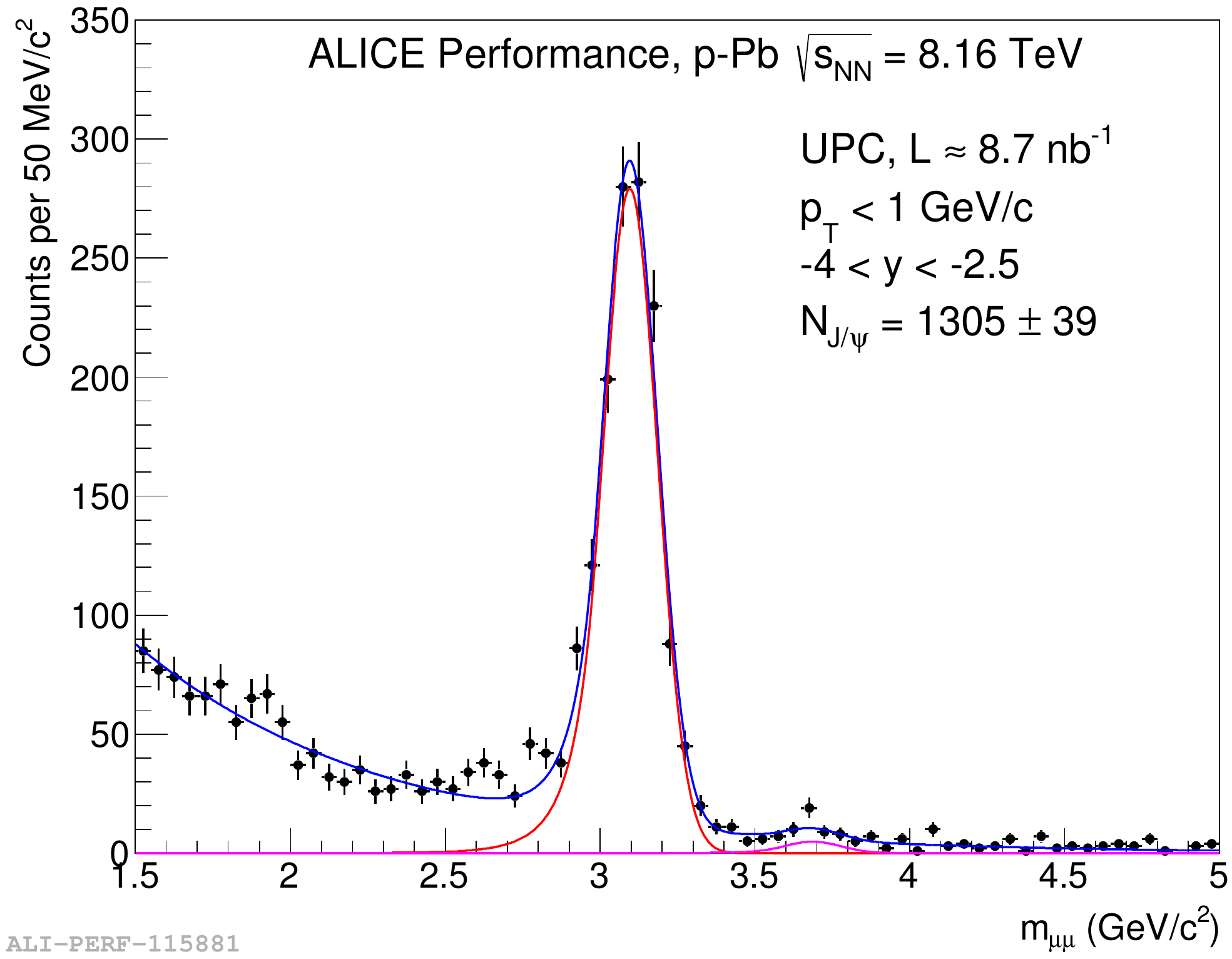}
\ \ \ \
\includegraphics[width=0.38\linewidth]{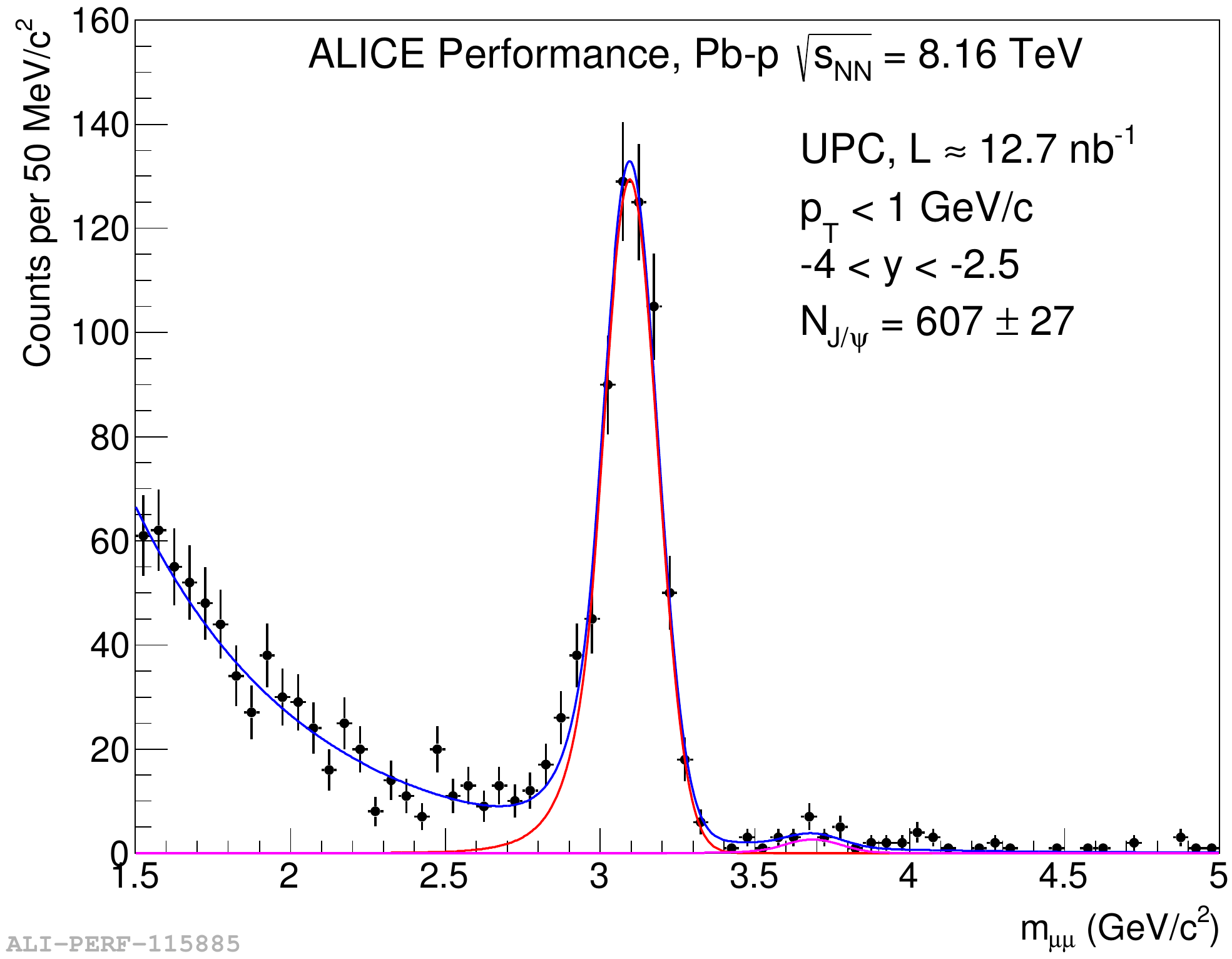}
\caption{
\label{fig:5}
Invariant mass distributions for unlike-sign dimuons with pair $p_{\rm T} < 1 {\rm\  GeV}/c$ and rapidity $-4 <y <-2.5$  in ultra-peripheral p-Pb (left) and Pb-p (right) collisions at $\sqrt{s_{\rm NN}} = 8.16$ TeV, corresponding to $\gamma p$ center-of-mass energy ranges $27 <W_{\gamma p} < 57$~GeV and $700 <W_{\gamma p} < 1480$~GeV respectively. $J/\psi$ and $\psi(2S)$ peaks are fitted with Crystal Ball functions on top of dimuon continuum fitted with an exponential function.
}
\vspace*{-0.5cm}
\end{figure}

Quarkonium photoproduction off protons in p--Pb UPC can be used to probe the behaviour of the gluon density at low $x$. ALICE has previously published results on the exclusive J$/\psi$ photoproduction off protons in p--Pb UPC  at~$\sqrt{s_{\mathrm{NN}}} = 5.02\ {\rm TeV}$~\cite{TheALICE:2014dwa}.
New data in p-Pb UPC collisions at $\sqrt{s_{\rm NN}} = 8.16$ TeV were collected in 2016, as shown in Fig.~\ref{fig:5}. This sample will provide further constraints on the $J/\psi$ photoproduction off protons and will probe the behaviour of the gluon density in the proton down to~$x\sim10^{-5}$.

\bibliographystyle{elsarticle-num}
\bibliography{kryshen}

\end{document}